\documentclass[12pt]{article}
\usepackage{amsmath}
\usepackage{pstricks}
\usepackage{amsthm}
\usepackage{youngtab}
\usepackage{bbm}
\usepackage{slashed}
\usepackage{amssymb}
\def\half{\frac{1}{2}}

\newfont{\bbbold}{msbm10 scaled \magstep1}

\def\bbC{\mbox{\bbbold C}}

\def\bbF{\mbox{\bbbold F}}

\def\bbR{\mbox{\bbbold R}}

\def\cA{{\cal A}}

\def\cD{{\cal D}}

\def\cF{{\cal F}}

\def\cL{{\cal L}}

\def\cN{{\cal N}}
\def\cO{{\cal O}}

\def\cR{{\cal R}}
\def\cS{{\cal S}}
\def\cT{{\cal T}}

\newfont{\goth}{eufm10 scaled \magstep1}

\def\gc{\mbox{\goth c}}

\def\gg{\mbox{\goth g}}
\def\gh{\mbox{\goth h}}
\def\gi{\mbox{\goth i}}

\def\gn{\mbox{\goth n}}
\def\go{\mbox{\goth o}}
\def\gp{\mbox{\goth p}}

\def\gs{\mbox{\goth s}}

\def\gu{\mbox{\goth u}}

\def\a{\alpha}\def\adt{\dot \alpha}
\def\b{\beta}\def\bdt{\dot \beta}
\def\c{\gamma}\def\C{\Gamma}\def\cdt{\dot\gamma}
\def\d{\delta}\def\D{\Delta}
\def\e{\epsilon}\def\ve{\varepsilon}
\def\F{\Phi}\def\vf{\varphi}
\def\h{\eta}

\def\k{\kappa}
\def\l{\lambda}\def\L{\Lambda}

\def\r{\rho}
\def\S{\Sigma}

\def\th{\theta}

\def\be{\begin{equation}}\def\ee{\end{equation}}
\def\bea{\begin{eqnarray}}\def\eea{\end{eqnarray}}
\def\barr{\begin{array}}\def\earr{\end{array}}

\def\o{\omega}\def\O{\Omega}
\def\U{\Upsilon}
\def\del{\partial}
\def\ua{\underline{\alpha}}
\def\ub{\underline{\phantom{\alpha}}\!\!\!\beta}
\def\uc{\underline{\phantom{\alpha}}\!\!\!\gamma}

\def\xz{\times}

\let\la=\label

\def\nn{\nonumber}
\def\bd{\begin{document}}
\def\ed{\end{document}}
\def\ba{\begin{array}}
\def\ea{\end{array}}
\def\bea{\begin{eqnarray}}
\def\eea{\end{eqnarray}}
\def\ft#1#2{\tfrac{#1}{#2}}
\def\fft#1#2{\frac{#1}{#2}}
\def\sst#1{{\scriptscriptstyle #1}}
\def\oneone{\rlap 1\mkern4mu{\rm l}}

\newcommand{\eq}[1]{(\ref{#1})}
\newcommand{\w}[1]{\\[0.#1cm]}
\def\eqs#1#2{(\ref{#1}-\ref{#2})}
\def\det{{\rm det\,}}
\def\tr{{\rm tr}}
\def\ad{{\rm ad}}

\newcommand{\hoch}[1]{$\, ^{#1}$}
\newcommand{\imperial}{\it\small Theoretical Physics Group, Imperial College London\\ Prince Consort Road, London SW7 2AZ, UK}
\newcommand{\kings}
{\it\small Department of Mathematics, King's College, University of London\\ Strand, London WC2R 2LS, UK}
\newcommand{\uu}
{\it\small Department of Theoretical Physics, Uppsala, Sweden}
\newcommand{\hip}
{\it\small HIP-Helsinki Institute of Physics, P.O. Box 64 FIN-00014
University of Helsinki, Suomi-Finland}
\newcommand{\stock}
{\it\small Department of Theoretical Physics, Stockholm, Sweden}
\newcommand{\golm}
{\it\small AEI, Max Planck Institut f\"ur Gravitationsphysik\\ Am M\"{u}hlenberg 1, D-14476 Potsdam, Germany}
\makeatletter
\renewcommand\theequation{\thesection.\arabic{equation}}
\@addtoreset{equation}{section} \makeatother

\newcommand{\sa}{/ \hspace{-1.2ex}}
\newcommand{\saa}{/ \hspace{-1.4ex}}
\newcommand{\saaa}{\, / \hspace{-1.6ex}}
\newcommand{\Scal}[1]{\Bigl ({#1} \Bigr )}
\newcommand{\scal}[1]{\bigl ({#1} \bigr )}

\newcommand{\CR}{\nonumber \\*}

\newcommand{\trace}{\hbox {tr}~}
\newcommand{\traceS}{\hbox {tr}_{\scriptscriptstyle \mathfrak{S}}~}

\DeclareMathAlphabet{\mathpzc}{OT1}{pzc}{m}{it}
\def\BRST{\,\mathpzc{s}\,}
\def\aBRST{{\scriptstyle (\mathpzc{s})}}
\def\q{{{\scriptscriptstyle (Q)}}}
\def\qs{{\scriptscriptstyle (Q\mathpzc{s})}}
\def\Qsla{{\mathcal{S}_{\q}}}
\def\Slav{{\mathcal{S}_\aBRST}}
\def\epsilonb{{\overline{\epsilon}}}
\def\bulletup{{\scriptstyle \bullet}}

\newcommand{\gra}[2]{{\scriptscriptstyle (#1 , #2 )}}
\newcommand{\ord}[1]{{\scriptscriptstyle (#1)}}

\def\cL{{\cal L}}
\def\cN{\mathcal{N}}
\def\cO{\mathcal{O}}

\def\ie{{\it i.e.}\ }
\def\eg{{\it e.g.}\ }

\newcommand{\sfrac}[2]{{\scriptstyle \frac{#1}{#2}}}
\newcommand{\stfrac}[2]{{\scriptscriptstyle \frac{#1}{#2}}}

 \def\balpha{{\overline{\alpha}}}
 \def\bbeta{{\overline{\beta}}}
 \def\bgamma{{\overline{\gamma}}}
 \def\bdelta{{\overline{\delta}}}
 \def\bepsilon{{\overline{\epsilon}}}
 \def\bvarepsilon{{\overline{\varepsilon}}}
 \def\bzeta{{\overline{\zeta}}}
 \def\bareta{{\overline{\eta}}}
 \def\btheta{{\overline{\theta}}}
 \def\bvartheta{{\overline{\vartheta}}}
 \def\biota{{\overline{\iota}}}
 \def\bkappa{{\overline{\kappa}}}
 \def\blambda{{\overline{\lambda}}}
 \def\bmu{{\overline{\mu}}}
 \def\bnu{{\overline{\nu}}}
 \def\bxi{{\overline{\xi}}}
 \def\bpi{{\overline{\pi}}}
 \def\brho{{\overline{\rho}}}
 \def\bvarrho{{\overline{\varrho}}}
 \def\bsigma{{\overline{\sigma}}}
 \def\bvarsigma{{\overline{\varsigma}}}
 \def\btau{{\overline{\tau}}}
 \def\bphi{{\overline{\phi}}}
 \def\bvarphi{{\overline{\varphi}}}
 \def\bchi{{\overline{\chi}}}
 \def\bpsi{{\overline{\psi}}}
 \def\bomega{{\overline{\omega}}}

\def\thalf{{\textrm{\tiny\textonehalf}}}
\def\tquarter{{\textrm{\tiny\textonequarter}}}
\def\Ko{{\scriptscriptstyle K}}
\def\tKo{\scriptscriptstyle k }
\def\corr{$\clubsuit$}

\newcommand{\auth}{\large P.S.\ Howe${}^{a,}$\footnote{email: paul.howe@kcl.ac.uk} and U. Lindstr\"om${}^{b,c,}$\footnote{email: ulf.lindstrom@physics.uu.se}}

\begin{document}

\renewcommand{\thefootnote}{\fnsymbol{footnote}}

\null
\begin{flushright}
{\small UUITP-28/20}\\
{\small Imperial-TP-2020-UL-03}\\
\vskip 1.5 cm
\end{flushright}

\begin{center}
{\Large{\bf Local supertwistors and conformal supergravity in six dimensions}}
\vspace{.75cm}

\auth
\end{center}
\vspace{.5cm}

\centerline{${}^a${\it \small Department of Mathematics, King's College London}}
\centerline{{\it \small The Strand, London WC2R 2LS, UK}}
\vspace{.5cm}
\centerline{${}^b${\it \small Department of Physics and Astronomy, Theoretical Physics, Uppsala University}}
\centerline{{\it \small SE-751 20 Uppsala, Sweden }}
\vspace{.5cm}
\centerline{${}^c${\it \small Theoretical Physics, Imperial College, London}}
\centerline{{\it \small Prince Consort Road, London SW7 2AZ, UK}}

\vspace{1cm}


\centerline{{\bf Abstract}}
\vskip .5cm
The local supertwistor formalism, which involves a superconformal connection acting on the bundle of such objects over superspace, is used to investigate
superconformal geometry in six dimensions. The geometry corresponding to $(1,0)$ and $(2,0)$ off-shell conformal supergravity multiplets, as well the associated finite super-Weyl 
transformations, are derived.
\vspace{2cm}

\leftline{Contribution to the Royal Society volume in honour of Michael Duff's 70th birthday}

\vspace{1cm}

\renewcommand{\thefootnote}{\arabic{footnote}}
\setcounter{footnote}{0}

\pagebreak
\tableofcontents
\setcounter{page}{1}


\section{Introduction}
Conformal symmetry has been extensively studied over the years because of its relevance to various aspects of theoretical physics: two-dimensional conformal theory models and statistical mechanics; four-dimensional $N=2$ and $4$ superconformal field theories; as an underlying symmetry that may be broken to Poincar\'e  symmetry in four-dimensional spacetime, and as  a tool to construct off-shell supergravity theories. It has repeatedly captured Professor Duff's attention over the years, not least in the context of anomalies in gravity and supergravity \cite{Capper:1975ig}-\cite{Borsten:2018jjm}.

Here we shall be interested in the geometry of local (super) conformal theories, in particular that of the $(2,0)$ conformal supergravity in six dimensions. In components, this supergravity was discussed twenty years ago in \cite{Bergshoeff:1999db}.

The first paper on $D=4, N=1$ conformal supergravity (CSG) \cite{Kaku:1978nz} used a formalism in which the entire superconformal group was gauged in a spacetime context. Although this was not a fully geometrical
set-up because supersymmetry does not act on spacetime itself, but rather on the component fields, this paper nevertheless introduced the idea that gauging conformal boosts and
scale transformations could be very useful. After Poincar\'e supergravity had been constructed in conventional (i.e. Salam-Strathdee \cite{Salam:1974jj}) superspace \cite{Wess:1977fn},\cite{Siegel:1978mj}, it was subsequently shown how scale transformations could be incorporated as super-Weyl transformations \cite{Siegel:1978mj,Howe:1978km}. On the other hand, in the completely different approach to superspace supergravity of \cite{Ogievetsky:1979ay}, super scale transformations were built in right from the start. However, it has turned out to be difficult to extend the latter approach to other cases, such as higher dimensions or higher $N$. 
The superspace geometry corresponding to all $D=4$ off-shell CSG multiplets was given in \cite{Howe:1980sy}
 using conventional superspace with an $SL(2,\bbC) \xz U(N)$ group in the tangent spaces together with real super-Weyl transformations. The superspace geometries corresponding to most other off-shell CSG multiplets have also been described, from the conventional point of view in $D=3$ \cite{Howe:1995zm,Kuzenko:2011xg,Gran:2012mg} and from conformal superspace in $D=3 ,4$ and $5$  \cite{3to5}, and in $D=6$ (the $(1,0)$ theory) \cite{Butter:2016qkx}, \cite{Butter:2017jqu}, \cite{Butter:2018wss}. The $D=6$ $(1,0)$ theory was also discussed earlier in harmonic superspace in \cite{Sokatchev:1988nk} and, some years ago, in projective superspace \cite{Linch:2012zh}.
    
In the non-supersymmetric case a standard approach is to work with conventional Riemannian geometry augmented by Weyl transformations of the metric. The Riemann tensor splits in two parts, the conformal Weyl tensor, and the Schouten tensor which is a particular linear combination of the Ricci tensor and the curvature scalar. This object transforms in a connection-type way under Weyl transformations and can be used to construct a new connection known as the tractor connection \cite{Thomas, Rocky, Gover:2008sw}. This takes its values in a parabolic subalgebra of the conformal algebra and acts naturally on a vector bundle whose fibres are $\bbR^{1,n+1}$ (in the Euclidean case), thus generalising in some sense the standard conformal embedding of flat $n$-dimensional Euclidean space. Although this idea does not carry over straightforwardly to the supersymmetric case, a similar construction, which does, can be made by replacing the $(n+2)$-dimensional fibre by the relevant twistor space. This formalism is called the local twistor formalism and was introduced in $D=4$ in \cite{Penrose:1986ca}. It has been discussed in the supersymmetric case for $N=1,2$ in $D=4$ by Merkulov, \cite{Merkulov:1986yr, Merkulov:1988zb, Merkulov:1989im}. Such a formalism depends on the dimension of spacetime because the twistor spaces also change.  In this paper we focus on the $D=6$ case for two reasons: firstly, local twistors in $D=6$ have not been discussed (to our knowledge) and secondly, we can use it to derive the standard superspace geometries corresponding to the off-shell $D=6$ CSG theories for which only the $(1,0)$ case has been given hitherto. 

In the following we shall take a slightly different approach to that of Merkulov in that we start from a connection taking its values in the full superconformal algebra. The conformal superspace formalism alluded to previously is a supersymmetric version of the Cartan connection formalism \cite{cartan} (first mentioned in the superspace context in \cite{Lott:2001st}). The formalism we advocate here can be thought of as an associated Cartan formalism in that the connection acts on a vector bundle rather than a principal one. The contents of the paper is as follows: in section 2 we briefly review conventional conformal geometry and the tractor formalism; in section 3 we  introduce local twistors in $D=6$, define connections and curvatures taking their values in the conformal Lie algebra in the twistor representation and rederive the results of section 2 from this point of view; in section 4 we discuss the local supertwistor formalism for both CSG theories in $D=6$; in section 5, we use this formalism to derive finite super-Weyl transformations in conventional superspace; in section 6 we discuss the constraints that are necessary to express the torsion and curvature components in terms of the irreducible conformal supergravity  multiplets and in section 7 we rederive the results of section 6 from a minimal approach in which the only constraint is that the dimension-zero torsion takes its flat form. In section 8 we summarise our results. There is also an appendix reviewing superconformal multiplets in $D=6$.
\section{Conformal geometry}

From a mathematical point of view a conformal structure in $n$ dimensions can be thought of as $G$-structure with $G=CO(1,n-1)$, the Lorentz group augmented by scale transformations, see, for example \cite{Rocky, Gover:2008sw}. This group does not preserve a particular tensor, but only a Lorentzian metric up to a scale transformation, so that angles but not lengths are invariant. Picking a given representative metric $g$ and with the usual choice of torsion-free connection, we can decompose the curvature as

 \be
R_{ab,}{}^{cd}=C_{ab,}{}^{cd} + 4\d_{[a}{}^{[c} P_{b]}{}^{d]}
\la{2.1}
\ee
where the trace-free part $C_{ab,cd}$ is the Weyl tensor and where
\be
P_{ab}= \left(R_{ab} -\frac{1}{2(n-1)} \h_{ab} R\right)\ 
\la{2.2}
\ee
is the Schouten tensor. Here, $\h_{ab}$ denotes the Lorentz metric with respect to an orthonormal frame, $R_{ab}$ is the usual symmetric Ricci tensor and $R$ the curvature scalar. Under a finite Weyl transformation of the metric $g\mapsto S^2 g$, we find
\be
\D P_{ab}=S^{-2}\left(D_a Y_b- Y_a Y_b +\half\h_{ab} Y^2\right)\ ,
\la{2.3}
\ee
where $\D$ denotes a finite change, and where $Y_a=S^{-1} D_a S$.  The prefactor $S^{-2}$ rises because we are working in an orthonormal frame.

In the conventional formalism one can define a new connection, the tractor connection, which takes its values in the Lie algebra of the conformal algebra $\gs\go (2,n)$, and which acts naturally to give a covariant derivative acting on vector fields taking their values in $\bbR^{2,n}$, the Schouten tensor being a key component in this connection. 
However, the tractor formalism cannot be  adapted directly to the supersymmetric case because the superconformal groups are not simply given by  super Lorentz groups in two higher dimensions, one of which is timelike.  Instead, one should think about supertwistors because they naturally carry the fundamental representations of superconformal algebras. It is therefore more relevant to study local twistor connections, introduced in \cite{Penrose:1986ca} in the non-supersymmetric case in four-dimensional spacetime. From this point of view one has to consider different twistor spaces according to the dimensions of spacetime. 

Before moving on to describe the local twistor formalism we briefly review the theory of Cartan connections of which conformal gauging is an example, and which we will make use of in the twistor context later.
Let $H,G$ be Lie groups, $H\subset G$, with respective Lie algebras $\gh,\gg$, and  let $P$ be a principal $H$-bundle over a base manifold $M$. A Cartan connection on $P$ is a $\gg$-valued form $\o$ equivariant with respect to $H$, and such that $\forall X\in \gh$ $\o(X)=X$ and $\o$ gives an isomorphism from $T_p P$ to $\gg$, for any point $p\in P$.

A simple example is given by an $n$-dimensional manifold $M$ with $G=SO(n)\ltimes \bbR^n$, $H=SO(n)$. Then $\gg=\gg_{-1}\oplus\gh$, where $\gg_{-1}$ corresponds to translations and $\gh$ to rotations. The translational part of $\o$ is identified with the soldering form, \ie the vielbein, while the $\gh$-part corresponds to an $\gs\go(n)$ connection. In the conformal case $\gh=\gg_0\oplus\gg_1$ where $\gg_0=\gs\go(n)\oplus \bbR$, and $\gg_1=\bbR^n$. So $\gg_o$ corresponds to rotations and scale transformations while $\gg_1$ corresponds to conformal boosts. The grading of the Lie algebra then corresponds to the dilatational weights of the various components. The curvature of $\o$, $\cR=d\o+\o^2$, also has components corresponding to this grading, and it is straightforward to see that they correspond to the torsion, the curvature and scale curvature, and the conformal boost field strength respectively. 

\section{Local twistors in $D=6$}

In this section  the local twistor formalism in six-dimensional spacetime will be discussed. We shall consider a vector bundle over spacetime with twistor space fibres. The conformal group acts on it linearly and locally, so that we can keep manifest conformal covariance by introducing a suitable connection. 

The complex conjugate of a four-component $D=6$ spinor $u^\a$ is denoted $\bar{u}^{\adt}$ but this representation is equivalent to the undotted one as there is a matrix 
 $B_\a{}^{\adt}$ relating the two, $\bar u^{\adt}= \bar u^\a B_{\a}{}^{\adt}$. $B$ is unitary, $B^* B=1$, and satisfies $B\bar B=-1$.\footnote{We use the six-dimensional conventions of  \cite{Howe:1983fr}} Similar remarks hold for the inequivalent spinor representation denoted by a lower index, $v_\a$ say. So a twistor $z$ consists of a pair of 4-component spinors and can be written
\be
z=\left(\begin{array}{c} u^\a \\ v_{\a} \end{array}\right)
\la{3.1}
\ee
An element $g$ of the conformal group $O(8)$acts on a twistor $z$ by matrix multiplication, $z\mapsto g z$.\footnote{Strictly  Spin(8) in this context.} The orthogonality condition for the metric $K$  is
\be
g K g^t=K=\left(\begin{array}{cc} 0 & 1_4 \\  1_4 &0 \end{array}\right)
\la{3.2}
\ee
where $1_4$ denotes the unit $4\xz 4$ matrix and $^t$ denotes transpose. This holds in complex spacetime and the reality condition, which holds in real spacetime, is
\be
g R g^*=R:=\left(\begin{array}{cc} 0 & B^{-1} \\  B &0 \end{array}\right)\ .
\la{3.3}
\ee
For an element $h$ of the Lie algebra, 
\be
\left(\begin{array}{cc} a & b\\  c &d \end{array}\right)
\la{3.4}
\ee
orthogonality implies that $b$ and $c$ are skew-symmetric, while $d^t=-a$, so 28 components altogether as expected. When reality is imposed we find that
\begin{align}
(bB)&=-(bB)^*\nn\w1
(B^{-1}c)&=-(B^{-1}c)^*\nn\w1
d&=-Ba^*B^{-1}\ .
\la{3.5}
\end{align}
In the above the entries in $h$ involve only undotted indices, whereas in the last equation the $B$-matrices convert some of them to dotted ones. The index structure for $h$ is given by
\be
h=\left(\begin{array}{cc} a^\a{}_\b & b^{\a\b} \\  c_{\a\b} &d_\a{}^\b\end{array}\right)\ .
\la{3.6}
\ee
where $b$ and $c$ are skew symmetric and where the trace of $a$ is equal to the trace of $d$. So $h$ is an element the conformal algebra and we can immediately introduce a conformal connection taking its values in this algebra by introducing a one-form $\cA$ with the same index structure, thus
\be
\cal{A}=\left(\begin{array}{c c }
\hat\o^\a{}_{\b}\ &ie^{\a\b} \\
if_{\a\b} & \hat{\o}_{\a}{}^{\b}\\
\end{array}\right)\ .
\la{3.7}
\ee
Here, 
\begin{align}
\hat\o^\a{}_\b&=\o^\a{}_\b +\half\d^\a_\b \o_0\nn\w1
\hat\o_\a{}^\b&=\o_\a{}^\b-\half\d_\a{}^\b\o_0
\la{3.8}
\end{align} 
where the unhatted terms are the Lorentz connections in the spin representations while $\o_0$ is the scale, or dilation, connection. We have
\be
\o^\a{}_\b=\frac{1}{4} (\c^{ab})^\a{}_\b\, \o_{ab}=-\o_\b{}^\a=-\frac{1}{4} (\c^{ab})_\b{}^\a\o_{ab}\ ,
\la{3.9}
\ee
where $\o_{ab}$ is the Lorentz connection in the vector representation. In addition, antisymmetric bi-spinors are equivalent to vectors via relations of the form
\be
e^{\a\b}=\half e^a (\c_a)^{\a\b} \Leftrightarrow e^a=-\half e^{\a\b} (\c^a)_{\a\b}\ .
\la{3.15}
\ee

The factors of $i$ in \eq{3.7} have been introduced so that $e$ and $f$ are real. The one-forms $e$ and $f$ will be identified with the vielbein (soldering form)  and the conformal boost connection respectively. The soldering process allows one to identify diffeomorphisms of the base space with translational gauge transformations so that the latter can be ignored, although we shall retain $e$ in the conformal connection $\cA$.

The curvature is
\be
\cF=d\cA+ \cA^2\ =\left(\begin{array}{c c }
\hat\cR^\a{}_{\b}\ &i\cT^{\a\b} \\
i\cS_{\a\b} & \hat{\cR}_{\a}{}^{\b}\\
\end{array}\right)\ ,
\la{3.10}
\ee
where the components are given by
\begin{align}
\label{3.11}
\cT^{\a\b}&=\hat D e^{\a\b}\nn\w1
\cS_{\a\b}&=\hat D f_{\a\b}\nn \w1
\hat\cR^\a{}_{\b}&=\hat R^\a{}_\b - e^{\a\c} f_{\c\b}\ .
\end{align}
$\hat D$ is the covariant exterior derivative with respect to both the Lorentz and scale connections, $\hat R$ denotes the sum of the corresponding curvatures, while 
$\hat{\cR}^{\a}{}_{\b}$ on the left-hand side is the conformally covariant version which, when the scale curvature is set to zero, becomes the Weyl tensor.
$\cT$ and $\cS$ are the curvatures corresponding to translations and conformal boosts respectively, the former being the standard torsion modified by the scale connection. 

We now consider a conformal transformation $g(C,S)$ depending on a conformal boost parameter $C_a$ and a scale parameter $S$ given by
\be
g(C,S)=\left(\begin{array}{cc} S^{-\half}& 0\\  iS^{-\half}C &S^{\half} \end{array}\right)\ ,
\la{3.12}
\ee
where $C$ is skew-symmetric (using the fact that a vector index is equivalent to a skew-symmetric pair of spinor indices).  It  also obeys the reality condition
\be
(B^{-1}C)^*=B^{-1}C\ .
\la{3.12.1}
\ee
The transformation of $\cA$ is given by
\be
\cA \mapsto dg^{-1} g + g^{-1} \cA g
\la{3.13}
\ee
from which we find
\begin{align} 
\label{3.14}
e^{\a\b}&\mapsto S e^{\a\b}\nn\w1
\hat\o^\a{}_\b&\mapsto\hat\o^\a{}_\b-e^{\a\c}C_{\c\b}+\half \d ^\a_\b Y\nn\w1
f_{\a\b}&\mapsto S^{-1}\left(f_{\a\b}-\hat D C_{\a\b}+ C_{\a\c} e^{\c\d} C_{\d\b}\right)\ ,
\la{3.14}
\end{align}
where $Y=S^{-1} dS$. For the curvatures we find:
\begin{align}
\cT^{\a\b}&\mapsto S \cT^{\a\b}\nn\w1
\hat \cR^\a{}_\b&\mapsto \hat\cR^\a{}_\b-\cT^{\a\c} C_{\c\b}\nn\w1
\cS_{\a\b}&\mapsto S^{-1}\left(\cS_{\a\b}-2\hat\cR_{[\a{}}^\c C_{|\c|\b]}+ C_{\a\c} \cT^{\c\d} C_{\d\b}\right)\ .
\end{align}

In components the last equation in \eq{3.11} reads
\begin{align}
\label{3.15}
\cR_{0ab}&=R_{0ab} + 2f_{[ab]}\nn\w1
\cR_{ab}{}^{cd}&=R_{ab}{}^{cd}+2 \d_{[a}{}^{[c} f_{b]}{}^{d]}\ ,
\end{align}
while the transformation of the scale connection is
\be
\o_0\mapsto \o_0 -C + 2Y\ ,
\la{3.16}
\ee
with $C=e^a C_a$. 

As usual, we can choose the Lorentz connection such that the torsion is zero, and we can also use the freedom to add an antisymmetric  tensorial part to the conformal boost connection $f_{ab}$ ($f_b=e^a f_{ab}$) to set the scale curvature $\cR_{0ab}$ to zero, after which $R_{0ab}=d\o_{0ab}=-2 f_{[ab]}$. But now, from the trace part of the middle equation in (3.16), we can make a choice of conformal gauge (using $C_a$ ) to set $\o_0=0$, after which the remaining $f_{ab}$ is symmetric.  At this stage the Lorentz curvatures on the right are the usual torsion-free Lorentzian ones while there is a residual conformal boost freedom given in terms of $S$ by the one-form $Y$, $C_a=2Y_a$. In fact, we can see that the second equation in \eq{3.15}  is the same as \eq{2.1} if we identify $\cR_{ab,}{}^{cd}$ with the Weyl  tensor $C_{ab,}{}^{cd}$ and $f_{ab}$ with $-2P_{ab}$.
 Furthermore, the transformation (2.3) can be seen to be the same as the third equation in (3.16) after $C$ has been identified with $Y$.

\section{Local supertwistors in $D=6$}

A supertwistor in $D=6$ can be written in the form
\be
Z=\left(\begin{array}{c}
u^\a\\
v_{\a}\\
\l_i\\
\end{array}\right)\ .
\la{4.1}
\ee
where $i=1,\ldots 2N$ for $(N,0)$ supersymmetry, $N=1,2$. Here $(u,v)$ are commuting objects while $\l$ is odd. The superconformal group is $OSp(8|N)$ in complex superspace and preserves the orthosymplectic metric $K$, so for an element $g$ of the group we have
\be
g K g^{st}=K
=\left(\begin{array}{c c |c}
0&1_4&0\\
1_4 & 0& 0\\
\hline
0&0& J_0
\end{array}\right)\ .
\la{4.2}
\ee
where $^{st}$ denotes the supertranspose, which is the same as the ordinary transpose except for an additional minus sign for each element in the bottom left (odd) sector.
The $2N \xz 2N$ matrix $J_0$ is the $Sp(N)$ symplectic invariant. In real spacetime we need to impose the reality constraint
\be
g R g^*=R=\left(\begin{array}{c c |c}
0&B^{-1}&0\\
B & 0& 0\\
\hline
0&0& 1_{2N}
\end{array}\right)\ .
\la{4.3}
\ee
An element of the Lie superalgebra, $h$, has the form
\be
h=\left(\begin{array}{c c |c}
a^\a{}_\b& b^{\a\b}&\ve^{\a j}\\
c_{\a\b} & d_\a{}^\b& \vf_\a{}^j\\
\hline
\l_{i \b}&\r_i{}^{\b}& e_i{}^j
\end{array}\right)\
\la{4.4}
\ee
The orthosymplectic constraint implies that $b$ and $c$ are skew-symmetric and $d=-a^t$, as before, while
\be
e J_0=-J_0 e^t
\la{4.5}
\ee
In indices, setting $(J_0)_{ij}=\h_{ij}$, this implies\footnote{Indices are raised and lowered according to the rule: $X^i=\h^{ij} X_j\Leftrightarrow X_i=X^j\h_{ji}$ with 
$\h^{ik}\h_{jk}=\d_j{}^i$.}
\be
e_{ij}:=e_i{}^k \h_{kj}=e_{ji}\ .
\la{4.6}
\ee
For the odd components we have
\be
\r=J_0 \ve^t \qquad   \l=J_0\vf^t 
\la{4.7}
\ee
or, in indices,
\begin{align}
\r_i{}^\b&=\h_{ij}\ve^{\b j}\Rightarrow \r_i{}^\b=-\ve^\b{}_i\nn\w1
\l_{i\b}&=\h_{ij}\vf_\b{}^j \Rightarrow \l_{i\b} = -\vf_{\b i}
\la{4.8}
\end{align}
Next we need to impose reality in order to move to real superspace. This is done with equation (3.3) but this time with $R$ extended by the unit matrix in the odd-odd sector, as in \eq{4.3}.
The result of imposing $gRg^*=g$, at the Lie algebra level is that $a,b,c$ and $d$ obey the same conditions as in the bosonic case while $e$ satisfies 
\be
e=-e^*
\la{4.9}
\ee
For the independent odd components of $h$ we have:
\begin{align}
\bar\ve^{\adt}_i&=-\h_{ij} \ve^{\b j} B_\b{}^{\adt}\nn\w1
\bar\vf_{\adt}^i&=(B^{-1})_{\adt}{}^\b \h^{ij}\vf_{\b j}\ ,
\la{4.10}
\end{align}
These constraints simply mean that $\ve^{\a i}$ and $\vf_{\a i}$ are symplectic Majorana-Weyl spinors as one would expect. They are respectively the parameters for $Q$ and $S$ supersymmetry transformations.

The connection $\cA$ is
\be
\cA=\left(\begin{array}{c c |c}
\hat\O^\a{}_\b& iE^{\a\b}& E^{\a j}\\
iF_{\a\b} & \hat\O_\a{}^\b& F_\a{}^j\\
\hline
-F_{i \b}&-E_i{}^{\b}& \O_i{}^j
\end{array}\right)\ ,
\la{4.11}
\ee
where $E^A=(E^a,E^{\a i})$,  with $E^a=\half (\c^a)_{\a\b}E^{\a\b}$, will be identified with the even and odd super-vielbein one-forms of the underlying superspace, 
$F_A=(F_a,F_{\a i})$ is the connection for superconformal transformations, \ie $S$-supersymmetry and standard conformal transformations, $\hat\O^\a{}_\b$  ($\hat\O_\a{}^\b$) is the Lorentz plus scale connection and $\O_i{}^j$ the internal $\gs\gp(N)$ connection. On the bottom line, $F_{i\b}$ and $E_i{}^\b$ are transposes of $F_\a{}^j$ and $E^{\a j}$ with the internal index lowered by $(J_0)_{ij}=\h_{ij}$.
The curvature two-form, $\cF=d\cA+\cA^2$, has components  given in matrix form by:
\be
\cal{F}=\left(\begin{array}{c c |c}
\hat \cR^\a{}_{\b}\ &i\cT^{\a\b} & \cT^{\a j}\\
i \cS_{\a\b} & \hat{\cR}_{\a}{}^{\b}&\cS_{\a}{}^j\\
\hline
-\cS_{i\b}&-{\cT}_i{}^{\b}& \hat\cR_i{}^j
\end{array}\right)\ ,
\la{4.12}
\ee
where
\begin{align}
\cT^{\a\b}&=\hat D E^{\a\b}+iE^{\a k} E_k{}^\b\nn\w1
\cT^{\a j}&=\hat D E^{\a j} + iE^{\a \c}F _\c{}^{j} \nn\w1
\hat{\cR}^\a{}_\b&=\hat R^\a{}_\b - E^{\a\c} F_{\c\b} - E^{\a k} F_{k\b}\nn\w1
\cR_i{}^j&=R_i{}^j -F_{i\c} E^{\c j} -E_i^{\c}F_{\c}{}^j\nn\w1
\cS_{\a\b}&=\hat D F_{\a\b} +iF_{\a}{}^k F_{k\b}\nn\w1
\cS_{\a}{}^j&=\hat D F_{\a}{}^j +iF_{\a\c} E^{\c j}\ .
\label{4.13}
\end{align}
Here, $\hat D$ is the superspace covariant exterior derivative with respect to scale, Lorentz and internal symmetries, while the leading terms on the right, for the top four lines, are the standard superspace and torsion and curvature tensors for the corresponding connections (extended by the scale connection).

The Bianchi identity is 
\be
\cD\cF:= d\cF + [\cF,\cA]=0\ .
\la{4.13.1}
\ee
Written out in components this is, for the torsions,
\begin{align}
\hat D \cT^{\a\b}-2E^{[\a|\c|} \hat\cR_\c{}^{\b]}-2i E^{[\a |k|} \cT^{\b]}_{\ \ k}&=0\nn\w1
\hat D\cT^{\a j}+\hat \cR^\a{}_\c E^{\c j}-E^{\a k}\cR_k{}^j+i\cT^{\a\c} F_\c{}^j-iE^{\a\c} \cS_\c{}^j&=0\ ,
\label{4.13.1}
\end{align}
for the Lorentz, scale and $\gs\gp(N)$ curvatures,
\begin{align}
D \hat \cR^\a{}_{\b} -\cT^{\a\c} F_{\c\b} -\cT^{\a k} F_{\b k}+ E^{\a\c} \cS_{\c\b} + E^{\a k} \cS_{\b k}&=0\nn\w1
 D\cR_{ij}-2\cT^\c_{(i} F_{|\c|j)} +2 E^\c_{(i} \cS_{|\c| j)}&=0\ ,
 \la{4.13.3}
 \end{align}
 and for the superconformal curvatures,
 \begin{align}
 \hat D\cS_{\a\b} +2\hat\cR_{[\a}{}^\c F_{|\c|\b]} +2 i \cS_{[\a}{}^k F_{\b] k}=0\nn\w1
 \hat D\cS_\a{}^j +\hat\cR_\a{}^\c F_\c{}^j-F_\a{}^k \cR_\k{}^j +i\cS_{\a\c} E^{\c j}-iF_{\a\c} T^{\c j}&=0\ .
 \la{4.13.4}
 \end{align}

\section{Finite Super-Weyl transformations}


We shall now repeat the steps carried out in the non-supersymmetric case to reduce the conformal and superconformal boost parameters to derivatives of the scale parameter.
We introduce a group element $g(S,C,\C)$ where $S$ is a scale parameter and $\C_{\a}{}^{i}$ is an $S$-supersymmetry parameter. It is given by
\be
g=\left(\begin{array}{c c |c}
S^{-\half}& 0& 0\\
iS^{-\half} \tilde C& S^{\half}& \C\\
\hline
S^{-\half}J_0\C^t& 0& 1
\end{array}\right)
\la{4.14}
\ee
where the index structure is as above, in \eq{4.11} for example, where $J_0$ is the $Sp(N)$ invariant discussed previously, and where
\be
\tilde C +\tilde C^t +i \C J_0 \C^t=0\ .
\la{4.15}
\ee
If we write
\be
\tilde C=C-\frac{i}{2} \C J_0\C^t
\la{4.17}
\ee
then, from \eq{4.15}, $C$ is antisymmetric since $\C J_0 \C^t$ is symmetric.
Reality implies that
\begin{align}
C&=B C^* B\ ,\ 
\nn\w1
\bar\C&=-B^{-1}\C J_0\ .
\la{4.16}
\end{align}
Note also that the latter equation implies that $\C J_0\C=B(\C J_0\C)^*B$. 

Under such a transformation the components of $\cA$ transform as follows:

\begin{align}
\label{4.18}
E^{\a\b}&\mapsto S E^{\a\b}\nn\w1
E^{\a j}&\mapsto S^{\half}(E^{\a j} +i E^{\a\b} \C_\b{}^j)\nn\w1
F_{\a\b}&\mapsto S^{-1}\left(F_{\a\b} -\hat D C_{\a\b} -i(\hat D \C_{[\a}{}^k)\C_{\b]k} +2iF_{[\a}{}^k \C_{\b]k}+i\tilde C_{\c\a} E^{\c\d} \tilde C_{\d\b}-2\tilde C_{\c[\a} E^{\c k} \C_{\b]k}\right)\nn\w1
F_\a{}^j&\mapsto S^{-\half}\left(  F_\a{}^j-\hat D\C_\a{}^j  + i(E^{\b j} +iE^{\b\c} E_\c{}^j)\tilde C_{\b\a}-E^{\b k} \C_{\a k} \C_{\b}{}^j\right) \nn\w1
\hat\O^\a{}_\b&\mapsto\hat\O^\a{}_\b - E^{\a\c} \tilde C_{\c\b}  - E^{\a k} \C_{\b k}+\half \d^\a_\b Y\nn\w1
\O_i{}^j&\mapsto \O_i{}^j +\C_{\a i} E^{\a j} +\C_{\a}{}^j E^{\a}_{\ i} + i\C_{\a i} E^{\a\b} \C_\b{}^j \ .
\la{4.18}
\end{align}

The curvature transformations are obtained from those for the potentials by replacing the latter by the former in the equations above. In addition, for the superconformal curvatures, the derivative terms in \eq{4.18} must be replaced by curvature terms as follows:
\begin{align}
\hat D C_{\a\b}&\mapsto  2\hat R_{[\a}{}^\c C_{|\c|\b]}\nn\w1
\hat D \C_\a{}^j&\mapsto R_\a{}^\b \C_\b{}^j + \C_{\a}{}^k R_k{}^j         \ .
\la{4.18.1}
\end{align}

If we take the trace of the third equation in \eq{4.13} we find that
\be
2\cR_0=2R_0- E^A F_A\ ,
\la{4.19}
\ee
where we have defined the super-vector-valued one-form $F_A=(F_a,F_{\a i})$. By adjusting this potential we can choose $\cR_0=0$ so that the (graded) antisymmetric part of $F_{AB}$ is now proportional to $R_{0 AB}$. Taking the trace of the transformation of $\hat\O^\a{}_\b$ we find
\be
2\O_0\mapsto 2\O_0 - E^a C_a -E^{\a i} \C_{\a i} + 2Y
\la{4.20}
\ee
so that we can use the parameters $C_a$ and $\C_{\a i}$ to set $\O_0=0$. This leaves residual transformations determined by the scale parameter $S$, 
\be
C_A=2 Y_A= 2S^{-1} D_A S\ ,
\la{4.21}
\ee
where $C_A=(C_a,\C_{\a i})$. We shall take the components of $Y$ to be given by $Y_A=(Y_a, \U_{\a i})$ in order to clearly distinguish the even and odd components where necessary. A similar discussion for the $D=3$ case can be found in \cite{Butter:2013goa}.

To summarise, having made the above tensorial shifts of the conformal and super-conformal potentials along with the gauge choices which set $R_0=\cR_0=0$, we arrive at the result that the supergeometry is given by the above set of potentials and curvatures but where now the hats can be dropped (from the derivatives as well as the curvatures), and with finite super-Weyl transformations\footnote{Finite 6D $(1,0)$ super Weyl transformations derived in a different  context will also appear in a forthcoming publication \cite{forth}.}  given by equations \eq{4.18} with the parameters $C$ and $\C$ replaced by $2Y_a$ and $2\U_{\a i}$. In addition the components $F_{AB}$ of the super-vector-valued one-form $F_B$ are now graded symmetric, and this tensor can be defined as the super Schouten tensor. 

\section{Constraints}

The above discussion is completely general in the sense that the geometry of the underlying superspace is unconstrained. To make contact with the fields of the conformal supergravity multiplets we will have to impose constraints. In this section we do this from the point of view of the discussion of the previous section, assuming that the scale connection is set to zero and that the super-conformal parameters are determined in terms of the scale parameter.  In particular, this means that we can drop the hats from the curvatures and the covariant derivatives.

The transformations of the covariant torsions and curvatures under super-Weyl transformations are given by
\begin{align}
\cT^{\a\b}&\mapsto S\cT^{\a\b}\nn\w1
\cT^{\a j}&\mapsto S^{\half}(\cT^{\a j} +2 i \cT^{\a\b} \U_\b{}^j)\nn\w1
\cR^\a{}_\b&\mapsto \cR^\a{}_\b -2\cT^{\a\c}( Y_{\c\b} +\frac{i}{2}  \U_{\c k} \U_\b{}^k) - 2\cT^{\a k} \U_{\b k}\nn\w1
\cR_i{}^j&\mapsto \cR_i{}^j +\U_{\a i} \cT^{\a j} +\U_{\a}{}^j \cT^{\a i} +4 i\U_{\a i} \cT^{\a\b} \U_\b{}^j\nn\w1
\la{5.1}
\end{align}

The basic constraint that we shall choose is to set the even torsion two-form to zero,
\be
\cT^a=0\ ,
\la{4.13.5}
\ee
which is clearly covariant.
Using this, conventional constraints corresponding to connection choices (including superconformal ones) and the Bianchi identities, one finds that the covariant torsions (\ie the torsion components of $\cF$) are given by
\begin{align}
\cT_{\a i\,\b j}{}^{\c k}&=0\nn\w1
\cT_{a \b j}{}^{\c k}&=(\c_a)_{\b\d} G^{\c\d}_j{}^k:=(\c^{bc})_{\b}{}^{\c}G_{abc j}{}^k \nn\w1
\cT_{ab}{}^{\c k}&=\Psi_{ab}{}^{\c k}\ ,
\label{4.13.6}
\end{align}
where $G_{abc jk}$ is anti-self-dual on $abc$ (by its definition), anti-symmetric on $jk$ and symplectic-traceless on $jk$ for $N=2$, and where $\Psi_{ab}{}^{\c k}$ is the gamma-traceless gravitino field strength. For the curvature tensor components we find
\begin{align}
\cR_{\a i \b j,kl}&=0 \nn\w1
\cR_{\a i\b j,cd}&=4i(\c^a)_{\a\b} G_{acd ij} \nn\w1
\cR_{a\b j,cd}&= -\frac{i}{2}( \c_a\Psi_{bc}-\c_c\Psi_{ab} -\c_b\Psi_{ca})_{\b j}\nn\w1
\cR_{a\b j,kl}&=-8(\c_a \chi)_{\b (k,l)j}\nn\w1
\cR_{ab,cd}&=C_{ab,cd}\nn\w1
\cR_{ab,kl}&= F_{ab,kl}\ .
\label{4.13.7}
\end{align}
The dimension three-halves field $\chi^\a_{i,jk}$ is antisymmetric on $jk$  : it is a doublet for $N=1$ while for $N=2$ it is in the {\bf{16}} of $\gs\gp(2)$, \ie it is symplectic-traceless on any pair of indices. The graviton field-strength $C_{ab,cd}$ has the symmetries of the Weyl tensor, while $F_{ab,kl}$ is the $\gs\gp(N)$ field-strength tensor. We have thus located all of the components of the conformal supergravity field strength supermultiplets except for the dimension-two scalars. These fields are all higher-dimensional components of $G$. At dimension three-halves we have
\be
D_{\a i} G_{abc jk}=(\c_{[a} \xi_{bc]})_{\a i,jk }+ (\c_{abc} \chi)_{\a i,jk}
\la{4.13.7.1}
\ee
where
\be
\xi_{ab i,jk}^\a=\frac{3i}{2} \h_{i[j} \Psi_{ab k]}^{\ \ \a}  +\frac{3i}{8} \h_{jk} \Psi_{ab i}^{\ \ \a}\ .
\la{4.13.8}
\ee

At dimension two, the Bianchi identity on the second line of  \eq{4.13.1} is
\be
D'_{\a i} \Psi_{bc}{}^{\d l}=Y_{bc,\a i,}{}^{\d l}+ i (S_{b\a i,\c}{}^l (\c_{c})^{\c\d} -(b\leftrightarrow c))
\la{4.13.10}
\ee
where $S_{b\a i,\c}{}^l $ is the dimension-two component of the S-supersymmetry curvature, and where
\be
Y_{bc,\a i}{}^{\d l}:=\cR_{bc,\a i}{}^{\d l} -2 D_{[b} \cT_{c]\a i}{}^{\d l} -  2\cT_{[b |\a i|}{}^{\e m} \cT_{c] \e m}{}^{\d l}\ .
\la{4.13.11}
\ee

The covariant derivative here, $D'$, is augmented by an off-diagonal-term in the connection involving the S-supersymmetry connection:
\begin{align}
D' X_a&=  D X_a -\frac{i}{2} (\c_a)^{\a\b} F_{\a}{}^j  X_{\b j}\nn\w1
D' X_{\a i} &=D X_{\a i}
\la{4.13.12}
\end{align}
Suppressing the spinor and internal indices on the objects occurring in \eq{4.13.11}, we find
\begin{align}
S_b=&\frac{i}{2} Y_{bc} \c^c +\frac{i}{20} Y_{cd} \c^{cd} \c_{b}\nn\w1
D' \Psi_{bc}&=Y_{bc} -\half Y_{[b|d|}\c^d{}_{c]}-\frac{1}{20} Y_{de} \c^{de} \c_{bc}\ .
\la{4.13.12.1}
\end{align}

Finally, the dimension-two scalar $C_{ij,kl}$ is given by 
\be 
C_{ij,kl}=D_{\a i} \chi^\a_{j,kl}\ .
\la{4.13.13}
\ee

It is straightforward to show that this field is in the $\bf{16}$ of $\gs\gp(2)$ for $N=2$, whereas it is a singlet for $N=1$.

We can now rewrite these constraints in terms of conventional superspace using \eq{4.13}. The basic constraint \eq{4.13.5} implies that the dimension-zero torsion takes its usual flat form
\be
T_{\a i\,\b j}{}^c=-i\h_{ij} (\c^c)_{\a\b}\ ,
\la{5.2.1}
\ee
while 
\be
T_{\a i \,b}{}^c=T_{ab}{}^c=0\ .
\la{5.3}
\ee
For the remaining torsion components we find
\be
T_{\a i\,\b j}{}^{\c k}=0\ ,
\la{5.3.1}
\ee
while at dimension one,
\be
T_{a \b j}{}^{\c k}=(\c_a)_{\b\d} G^{\c\d}_j{}^k + \frac{i}{2} (\c_a)^{\c\d} F_{\b j,\d{}}^{\ \ \ \ k}\ ,
\la{5.3.2}
\ee
and at dimension three-halves,
\be
T_{ab}{}^{\c k}=\Psi_{ab}{}^{\c k}+ (\c_{[a})^{\c \d} F_{b] \d}{}^k .
\la{5.4}
\ee
where the first term is the gravitino field strength as remarked above. So the conventional superspace torsions differ from the covariant ones by by components of the super Schouten tensor $F_{AB}$. The same is true for the curvature components:
\begin{align}
R_{\a i\b j,cd}&=4i(\c^a)_{\a\b} G_{acd ij} -\frac{1}{2}( F_{\a i,\d j}(\c_{cd})^\d{}_\b +F_{\b j,\d i}(\c_{cd})^\d{}_\a)\ ,\nn\w1
R_{a\b j,cd}&= -\frac{i}{2}( \c_a\Psi_{bc}-\c_c\Psi_{ab} -\c_b\Psi_{ca})_{\b j} -\half F_{a \c j} (\c_{cd})^\c{}_\b -\h_{a[c} F_{|\b j |d]}\ ,\nn\w1
R_{ab,cd}&=C_{ab,cd} -2 \h_{[a [c} F_{b] d]}\ ,\nn\w1
R_{\a i \b j,kl}&=\h_{i(k} F_{\b j,\a l)} +\h_{j(k} F_{\a i,\b l)}\ ,\nn\w1
R_{a\b j,kl}&=-8(\c_a \chi)_{\b (k,l)j}+2 \h_{j(k} F_{a \b l)}\ ,\nn\w1
R_{ab,kl}&= F_{ab,kl}\ .
\label{5.4.1}
\end{align}
The standard superspace torsion and curvature components \eq{5.2.1} to \eq{5.4.1} therefore differ from the covariant ones \eq{4.13.6} and \eq{4.13.7}, which involve only the fields of the conformal supergravity multiplets \eq{6.7}, by various components of the super Schouten tensor.

\section{Minimal approach}

We shall now describe the superspace geometry corresponding to these conformal supergravity multiplets from a minimal perspective. We define a superconformal structure on a supermanifold with $(\rm{even}| \rm{odd})$ dimension $(6|8N)$ to be a choice of odd tangent bundle $T_1$ (of dimension $(0|8N)$ which is maximally non-integrable, so that the even tangent bundle $T_0$ is generated by commutators of sections of $T_1$, and such that the Frobenius tensor, $\bbF$, defined below, is invariant under $\bbR\oplus spin(1,5)\oplus\gs\gp(N)$. The components of $\bbF$ with respect to local bases $E_{\a i}$ for $T_1$ and $E^a$ for $T_0^*$ are given by
\be
\bbF_{\a i \b j}{}^c= \langle[E_{\a i}, E_{\b j}], E^c\rangle=-i\h_{ij}(\c^c)_{\a\b}\ ,
\la{6.8}
\ee
where $\langle,\rangle$ denotes the pairing between vectors and forms. The  $\bbR$ factor denotes an infinitesimal scale transformation, $\d E^a=SE^a$, $\d E_{\a i}=-\half  S E_{\a i}$, while the spin and symplectic algebras act in the natural way on the spacetime and internal indices. 

We now introduce connections for $\gs\gp(N)$ and $spin(1,5)$ and define the torsion and curvatures in the usual way. Note that this procedure involves the complementary basis $E_{\a i}$ for $T_1^*$ which is only determined modulo $T_0^*$, i.e. shifts of the form
\be
E_{\a i}\mapsto E_{\a i} + L_{\a i}{}^b E_b\ .
\la{6.9}
\ee
We could include this in the structure group, along with a corresponding connection, but we shall instead follow the standard procedure of using this freedom to impose some additional constraints at dimension one-half. In addition we shall not include a scale connection so that we have the standard superspace geometrical set-up. 

Identifying $\bbF_{\a i \b j}{}^c$ with the dimension-zero torsion $T_{\a i \b j}{}^c$, imposing suitable constraints on various components of the torsion corresponding to fixing the odd basis $E_{\a i}$ using \eq{6.9} and making appropriate choices for the $spin(1,5)$ and $\gs\gp(N)$ connections, one can show, with the aid of the usual superspace Bianchi identities and some algebra, that the components of the torsion and curvature tensors can be chosen to agree with those listed in equations \eq{5.2.1} to \eq{5.4.1}.

Under finite super-Weyl transformations we have

\begin{align}
\label{6.10}
E^{\a\b}&\mapsto S E^{\a\b}\nn\w1
E^{\a j}&\mapsto S^{\half}(E^{\a j} +2i E^{\a\b} \U_\b{}^j)\nn\w1
\O^\a{}_\b&\mapsto \O^\a{}_\b - E^{\a\c} \tilde C_{\c\b}  - E^{\a k} \U_{\b k}+\half \d^\a_\b Y\nn\w1
\O_{ij}&\mapsto \O_{ij} - 4E^\a_{(i}\U_{\a j)} + 4i\U_{\a i} E^{\a\b} \U_{\b j}\ .
\end{align}

with
\be
\tilde C_{\a\b}=C_{\a\b} -2i \U_{\a k} \U_\b{}^k\ ,
\label{6.11}
\ee

where 

\be 
C_A=2Y_A=2(Y_a,\U_{\a i})=2S^{-1} D_A S\ ,
\la{6.12}
\ee

with $Y_A$ being the components of the one-form $Y$, as before in \eq{4.21}.

In addition to the fields of the conformal supergravity multiplet, this geometry will also contain the components of the super Schouten tensor $F_{AB}$, whose transformations can be found in \eq{4.18}. We can recover the covariant forms for the torsions and curvatures by reversing the steps made earlier.

\section{Summary}

In this article we have discussed the supergeometries describing off-shell conformal supergravity multiplets in $D=6$ for $(N,0)$ supersymmetries with $N=1,2$ from the perspective of local supertwistors. In this formalism one introduces connections taking their values in the superconformal algebras in the twistor representation, which can be thought of as an associated version of the Cartan connection formalism. From this starting point one can then derive the standard superspace formalism in a systematic fashion. In order to specialise to the minimal off-shell conformal supergravity multiplets one then has to impose the constraint \eq{4.13.5}. We also showed that the same results can be obtained from the minimal formalism in which only the dimension-zero torsion, or Frobenius tensor, is specified.  This formalism has also been applied to $D=3$ \cite{Howe:1995zm,Gran:2012mg} while it
was shown in the  $D=4$ case that the super geometries also follow from the dimension-zero torsion constraint \cite{Howe:1980sy} superconformal geometries, although an additional constraint is required in the $N=4$ case.

\bigskip
\bigskip
\noindent{\bf\Large Acknowledgements}
\vskip1mm
\noindent 
We are grateful for communications with Sergei Kuzenko. U.L. also acknowledges  hospitality from the theory group at Imperial College, London, support from the EPSRC programme grant ``New Geometric Structures from String Theory'' EP/K034456/1 and support from Lars Hierta's foundation. 

\bigskip

\section{Appendix: Conformal supermultiplets}

In this section we shall briefly summarise the relevant conformal supermultiplets in $D=6$. \cite{Howe:1983fr,Koller:1982cs}

The basic superconformal matter multiplets are the tensor multiplets whose components include scalars, spinors and a two-form gauge field with a self-dual gauge field. For the $(1,0)$ case, the field strength superfield is a real scalar $\F$ satisfying the second-order constraint
\be
D_{\a (i} D_{\b j)}\F=0\ .
\la{6.1}
\ee
The independent components are $F, D_{\a i} \F$ and $\h^{ij} D_{\a i} D_{\b j}\F$ evaluated at the $\th=0$, corresponding to the scalar field, the fermion and the three-form field strength. These components are all on-shell. 
For the $(2,0)$ case the scalars are in the $5$ of $\gs\gp(2)$, and we have
\be
D_{\a i} \F_{jk}=\h_{ij}\l_{\a k}-\h_{ik}. \l_{\a j} +\frac{1}{2} \h_{jk} \l_{\a i}\ .
\la{6.2}
\ee
The only other independent component occurs at the next level in $\th$ and is again a three-form field strength. It is not difficult to verify that these multiplets can also be described in terms of a closed  super-three-form field strength $H$ where the lowest-dimensional non-zero component is
\begin{align}
H_{a\, \b j\,\c\,k}&= i(\c_a)_{\b\c} \h_{jk} \F,\ \ N=1\nn\w1
H_{a\, \b j\,\c\,k}&= i(\c_a)_{\b\c} \F_{jk}, \  \ \  \,N=2\ .
\la{6.3}
\end{align}
From these multiplets one can construct conformal supercurrent multiplets as bi-linears:
\begin{align}
J&=\F^2\ \ \qquad  \ \ N=1\nn\w1
J_{ij,kl}&= (\F_{ij} \F_{kl})_{\bf 14}\ N=2\ ,
\la{(6.4}
\end{align}
where the notation for $N=2$ indicates the projection onto $14$-dimensional representation (completely traceless with respect to the symplectic form $\h^{ij}$). The supercurrents obey the constraints
\begin{align}
(D^{3})^{\a}_{ijk}J&=0 \ \qquad  \qquad N=1\nn\w1
D_{\a m} J_{ij,kl}&= \h_{m[i}\chi_{\a j],kl} +\h_{m [k} \chi_{\a l],ij} \ N=2\ ,
\la{(6.4}
\end{align}
where $\chi_{\a i,jk}$ transforms under the $\bf{16}$-dimensional representation of $\gs\gp(2)$. Its totally antisymmetric part and symplectic traces all vanish. In the first line the cubic derivative is totally symmetric on $ijk$. 

The components of the current multiplets are
\begin{align}
J  &\rightarrow \L_i \rightarrow J_{a ij} + L_{abc}\rightarrow \S_a{}^i\rightarrow T_{ab} \nn\w1
J_{ij,kl} & \rightarrow \L_{i,jk} \rightarrow J_{a ij} + L_{abc, ij}\rightarrow \S_a{}^i\rightarrow T_{ab}
\la{6.5}
\end{align}
where the $J_{a ij}(=J_{a ji})$ are conserved currents for $\gs\gp(N)$, the $\S_a{}^i$ are the spin-three-halves currents, conserved and gamma-traceless, and $T_{ab}$ is the traceless conserved energy-momentum tensor. The antisymmetric $L$-tensors are self-dual, and in the $N=2$ case, antisymmetric and symplectic-traceless on the $ij$ indices. The total number of components are $40+40$ for $N=1$ and $128+128$ for $N=2$. 

The conformal supergravity multiplets are dual to the current multiplets. Their components are
\be
g_{mn}\rightarrow \psi_m{}^{}\rightarrow A_m{} + G_{abc} \rightarrow \chi \rightarrow C\ ,
\la{6.6}
\ee
where $g_{mn}$ is the metric, $\psi_m$ the gravitini, the $ A_m$ are the gauge fields for $\gs\gp(N)$ and $\chi$ and $E$ the dimension-three-halves fermions and the dimension-two scalars respectively. They are in the same representations of $\gs\gp(N)$ as  $\L$ and $J$. 

The field strengths for the supergravity multiplets are 
\be
G_{abc ij}\rightarrow \chi^{\a}_{i,jk} + \Psi_{ab}{}^{\a k} \rightarrow C_{ij,kl} + F_{ab,kl} + C_{ab,cd}\ .
\la{6.7}
\ee
where the leading $G$ components are now anti-self-dual with dimension one, in the $\bf{1}$ and $\bf{5}$ representations of $\gs\gp(N)$, $\Psi$ denotes the completely gamma-traceless gravitino field strengths, $E$ denotes the dimension-two scalars, in the $\bf{1}$ and $\bf{14}$ representations and $C$ is the Weyl tensor.

\end{document}